\documentclass[11pt]{article}
\usepackage{amssymb,amsmath,amsthm,latexsym}
\newcommand{\comment}[1]{} 

\begin{document}

\title{An Efficient Construction of Self-Dual Codes}

\author{
Yoonjin Lee\footnote{
The author was supported by Priority Research Centers Program
through the National Research Foundation of Korea(NRF) funded by the
Ministry of Education, Science and Technology(2009-0093827).} \\
Department of Mathematics\\
Ewha Womans University\\
11-1 Daehyun-Dong, Seodaemun-Gu, Seoul\\
120-750, S. Korea \\
e-mail: yoonjinl@ewha.ac.kr \\ \\
Jon-Lark Kim \footnote{Corresponding author}\\
Department of Mathematics\\
328 Natural Sciences Building \\
University of Louisville\\
Louisville KY 40292 USA \\
e-mail: jl.kim@louisville.edu\\
}

\date{Jan. 26, 2012}
\maketitle

\newtheorem{theorem}{Theorem}[section] 
\newtheorem{claim}[theorem]{Claim}
\newtheorem{lemma}[theorem]{Lemma}
\newtheorem{proposition}[theorem]{Proposition}
\newtheorem{corollary}[theorem]{Corollary}
\newtheorem{definition}[theorem]{Definition}
\newtheorem{remark}[theorem]{Remark}
\newtheorem{exercise}[theorem]{Exercise}
\newtheorem{example}[theorem]{Example}

\newcommand{\ob}{\mbox{$\overline{\omega}$}}
\newcommand{\om}{\mbox{$\omega$}}
\newcommand{\Tr}{\mbox{Tr}}
\newcommand{\C}{\mbox{$\cal C$}}
\newcommand{\CperpE}{\mbox{$\C^{\bot_E}$}}
\newcommand{\CperpH}{\mbox{$\C^{\bot_H}$}}
\newcommand{\br}{\mbox{$\bf{r}$}}
\newcommand{\bl}{\mbox{$\bf{l}$}}
\newcommand{\ba}{\mbox{$\bf{a}$}}
\newcommand{\be}{\mbox{$\bf{e}$}}
\newcommand{\bx}{\mbox{$\bf{x}$}}
\newcommand{\by}{\mbox{$\bf{y}$}}
\newcommand{\bv}{\mbox{$\bf{v}$}}
\newcommand{\bw}{\mbox{$\bf{w}$}}

\renewcommand{\a}{\mbox{$\alpha$}}
\renewcommand{\b}{\mbox{$\beta$}}
\newcommand{\GR}{\mbox{$\rm GR$}}
\newcommand{\R}{\mbox{${\rm GR}(p^m, r)$}}
\newcommand{\Rr}{\mbox{${\rm GR}(2^m, r)$}}
\newcommand{\F}{\mbox{$\mathbb F$}}
\newcommand{\Z}{\mbox{$\mathbb Z$}}

\font\msbm=msbm10 at 12pt
\newcommand{\ZZ}{\mbox{\msbm Z}}
\newcommand{\RR}{\mbox{\msbm R}}
\newcommand{\CC}{\mbox{\msbm C}}
\newcommand{\FF}{\mbox{\msbm F}}

\def\vx{\mathbf{x}}
\def\vy{\mathbf{y}}
\def\vv{\mathbf{v}}
\def\vu{\mathbf{u}}
\def\vw{\mathbf{w}}
\def\vc{\mathbf{c}}
\def\vd{\mathbf{d}}
\def\ma{\mathfrak{a}}

\begin{abstract}
We complete the building-up construction for self-dual codes by
resolving the open cases over $GF(q)$ with $q \equiv 3 \pmod 4$, and
over $\Z_{p^m}$ and Galois rings $\GR(p^m,r)$ with an odd prime $p$
satisfying $p \equiv 3 \pmod 4$ with $r$ odd. We also extend the
building-up construction for self-dual codes to finite chain rings.
Our building-up construction produces many new interesting self-dual
codes. In particular, we construct $945$ new extremal self-dual
ternary $[32,16,9]$ codes, each of which has a trivial automorphism
group. We also obtain many new self-dual codes over $\mathbb Z_9$ of
lengths $12, 16, 20$ all with minimum Hamming weight $6$, which is
the best possible minimum Hamming weight that free self-dual codes
over $\Z_9$ of these lengths can attain. From the constructed codes
over $\mathbb Z_9$, we reconstruct optimal Type I lattices of
dimensions $12, 16, 20,$ and $24$ using Construction $A$; this shows
that our building-up construction can make a good contribution for
finding optimal Type I lattices as well as self-dual codes. We also
find new optimal self-dual $[16,8,7]$ codes over $GF(7)$ and new
self-dual codes over $GF(7)$ with the best known parameters
$[24,12,9]$.
\end{abstract}

\section{Introduction}
\label{sec:intro}

Self-dual codes have been of great interest because they often
produce optimal codes and they also have beautiful connections to
other mathematical areas including unimodular lattices,
$t$-designs, Hadamard matrices, and quantum
codes (see~\cite{RaiSlo} for example).

There are several ways to construct self-dual codes. Early
constructions are based on gluing vectors, which work well when the
minimum distances of the codes are small~(cf. \cite{LeoPleSlo,
Ple75}). One powerful method is the balance principle~\cite{HufPle,
Koc89}, which restricts the generator matrix of a self-dual code.
Another general one is to build self-dual codes from self-dual codes
of smaller lengths. The first such construction is based on shadow
codes~\cite{BruPle, Dou95}. M. Harada~\cite{Har97} introduced an
easy way to generate many binary self-dual codes from a self-dual
code of a smaller length, and then later the first author~\cite{Kim}
introduced the so-called {\it building-up construction} for binary
self-dual codes. This construction says that any binary self-dual
code can be built from a self-dual code of smaller length. A few
years later, the building-up construction for self-dual codes over
finite fields $GF(q)$ was developed when $q$ is a power of $2$ or $q
\equiv 1 \pmod{4}$~\cite{KimLee}, and then over finite ring
$\Z_{p^m}$ with $p \equiv 1 \pmod{4}$~\cite{LeeLee}, and over Galois
rings $GR(p^m,r)$ with $p \equiv 1 \pmod{4}$ with any $r$ or $p
\equiv 3 \pmod{4}$ with $r$ even~\cite{KimLee2}, where $m$ is any
positive integer. It turns out that the building-up construction is
so efficient that one can easily find many (often new) self-dual
codes of reasonable lengths (e.g.,~\cite{GulKimLee}).

In this paper, we complete the open cases of the building-up
construction for self-dual codes over $GF(q)$ with $q \equiv 3 \pmod
4$, and over $\Z_{p^m}$ and Galois rings $GR(p^m,r)$ with an odd
prime $p$ such that $p \equiv 3 \pmod 4$ with $r$ odd. We also present a
building-up construction for self-dual codes over finite chain rings.

Our building-up construction yields many new interesting self-dual
codes. In fact, only one extremal self-dual ternary $[32,16,9]$ code
with a trivial automorphism group was known~\cite{Har_01} before. In
this paper, we construct $945$ new extremal self-dual ternary
$[32,16,9]$ codes, each of which has a trivial automorphism group,
i.e., the monomial group of order $2$.  We also obtain $208$ new optimal
self-dual $[16,8,7]$ codes over $GF(7)$ and $59$ new self-dual
codes over $GF(7)$ with the best known parameters $[24,12,9]$.
Furthermore, we construct many new self-dual codes over $\mathbb
Z_9$ of lengths $12, 16, 20$ all with minimum Hamming weight $6$,
which is the best possible minimum Hamming weight that free
self-dual codes over $\Z_9$ of these lengths can have. From the
self-dual codes over $\Z_9$ constructed by our building-up method,
we reconstruct optimal Type I lattices of dimensions $12, 16, 20,$
and $24$ using Construction $A$ (refer to~\cite{BanDouHarOur99,
ConSlo99, Gab04}). This shows that our building-up construction can
make a good contribution for finding optimal Type I lattices as well
as self-dual codes.

All our codes will be posted on {\tt www.math.louisville.edu/$\sim$jlkim/preprints}.

\comment{
This paper is organized as follows. We discuss the building-up
construction for self-dual codes over $GF(q)$ with $q \equiv  3
\pmod{4}$ in Section~\ref{sec:building-up}.
In Section~\ref{sec:building_rings}, we work on the building-up
method over finite chain rings, therefore over Galois rings
$GR(p^m,r)$ and finite rings $\Z_{p^m}$ with $p \equiv  3
\pmod{4}$ (since Galois rings and $\Z_{p^m}$ are special cases of
finite chain rings). We also discuss the building-up construction
over the $p$-adic rings for any prime $p>2$.
Examples of new extremal self-dual ternary $[32,16,9]$ codes are
given in Table~\ref{tab:F3} and examples of self-dual codes over
$\mathbb Z_9$ of lengths $12, 16$, and $20$ are described in
Tables~\ref{tab:Z9_1}-\ref{tab:Z9_3}. Our conclusion is given in
Section~\ref{sec:con-Building}.
}  

\section{Building-up construction for self-dual codes
over $GF(q)$ with $q \equiv  3 \pmod{4}$}
\label{sec:building-up}

In this section we provide the building-up construction for
self-dual codes over $GF(q)$ with $q \equiv 3 \pmod 4$, where $q$ is
a power of an odd prime. It is known~\cite[p. 193]{RaiSlo} that if
$q \equiv 3 \pmod 4$ then a self-dual code of length $n$ exists if
and only if $n$ is a multiple of $4$. Our building-up construction
needs the following known lemma~\cite[p. 281]{IreRos}.

\begin{lemma}
\label{lemma1} Let $q$ be a power of an odd prime with $q \equiv 3
\pmod 4$. Then there exist $\alpha$ and $\beta$ in $GF(q)^\ast$ such
that $\alpha^2+\beta^2+1=0$ in $GF(q)$, where $GF(q)^\ast$ denotes
the set of units of $GF(q)$.
\end{lemma}

We give the {\em building-up construction} below and prove that it
holds for any self-dual code over $GF(q)$ with $q \equiv 3 \pmod
4$.

\begin{proposition}
\label{prop:BU} Let $q$ be a power of an odd prime such that $q
\equiv 3 \pmod 4$, and let $n$ be even. Let $\alpha$ and $\beta$
be in $GF(q)^\ast$ such that $\alpha^2+\beta^2+1=0$ in $GF(q)$.
Let $G_0=(\br_i)$ be a generator matrix (not necessarily in standard form)
of a self-dual code $\C_0$ over $GF(q)$ of length $2n$, where $\br_i$ are
the row vectors for $1 \le i \le n$. Let $\mbox{\bf x}_1$ and
$\mbox{\bf x}_2$ be vectors in $GF(q)^{2n}$ such that $\mbox{\bf
x}_1\cdot\mbox{\bf x}_2 = 0$ in $GF(q)$ and $\mbox{\bf
x}_i\cdot\mbox{\bf x}_i = -1$ in $GF(q)$ for each $i=1,2$. For each
$i$, $ 1 \le i \le n$, let $s_i:= \mbox{\bf x}_1 \cdot \br_i$,
$t_i:= \mbox{\bf x}_2 \cdot \br_i$, and $\by_i := ( -s_i, -t_i,
-\a s_i-\b t_i, -\b s_i+\a t_i)$ be a vector of length $4$. Then
the following matrix

\[ G = \left[ \begin{array}{cccc|ccccccc}
 1 &0 &0 &0 & & &&\bx_1 &&& \\
 0 &1 &0 &0 & & &&\bx_2 &&& \\ \hline
 & \by_1 && & &  &&\br_1 &&& \\
 & \vdots && & &  && \vdots &&& \\
  & \by_n && & &  &&\br_n &&& \\
\end{array}
\right]
\]

generates a self-dual code \C\ over $GF(q)$ of length $2n+4$.
\end{proposition}

\begin{proof}
We first show that any two rows of $G$ are orthogonal to each other.
Each of the first two rows of $G$ is orthogonal to itself as the
inner product of the $i$th row with itself equals $1+ \mbox{\bf
x}_i\cdot\mbox{\bf x}_i= 0$ in $GF(q)$ for $i=1, 2$. The first row
of $G$ is orthogonal to the second row of $G$ as
$\mbox{\bf x}_1\cdot\mbox{\bf x}_2= 0$ in $GF(q)$. Furthermore, the
first row of $G$ is orthogonal to any $(i+2)$th row of $G$ for $1
\le i \le n$ since the inner product of the first row of $G$ with
the $(i+2)$th row of $G$ is
$$(1,0, 0, 0 ) \cdot \by_i + \bx_1 \cdot \br_i = -s_i + s_i =0.$$
Similarly, the second row of $G$ is orthogonal to any $(i+2)$th row of $G$
for $1 \le i \le n.$
We note that $\br_i \cdot \br_j =0$ for $1 \le i,j \le n.$
Any $(i+2)$th row of $G$ is orthogonal to any $(j+2)$th row for $1
\le i,j \le n$ because the inner product of the $(i+2)$th row of $G$
with the $(j+2)$th row is equal to
$${\by_i}\cdot{\by_j} + \br_i\cdot\br_j= (1+\a^2+\b^2)(s_is_j + t_it_j) = 0
\ \ \mbox{ in } GF(q).$$ Therefore, \C\ is self-orthogonal; so $\C \subseteq \C^\perp.$

We claim that the code \C\ is of dimension $n+2$. It suffices to show that
no nontrivial linear combination of the first two rows of $G$ is in the span of
the bottom $n$ rows of $G$. Assume such a combination exists.
Then $c_1({\mbox{the first row of $G$}}) + c_2({\mbox{the second row of $G$}}) =
\sum_{i=1}^{n} d_i(\by_i, \br_i)$ for some nonzero $c_1$ or $c_2$ in $GF(q)$
and some $d_i$ in $GF(q)$ with $i=1, \ldots, n$.
Then comparing the first four coordinates of the vectors in both sides, we get
$c_1 = - \sum_{i=1}^{n} d_i s_i,$ $c_2 = - \sum_{i=1}^{n} d_i t_i,$
$0 = - \sum_{i=1}^{n} d_i (\a s_i + \b t_i), \ 0 =  \sum_{i=1}^{n} d_i (-\b s_i +\a t_i)$;
thus $0 = - \sum_{i=1}^{n} d_i (\a s_i + \b t_i) =
\a ( -\sum_{i=1}^{n} d_i s_i ) + \b ( -\sum_{i=1}^{n} d_i t_i ) = \a c_1 +\b c_2$, that is,
we have $\a c_1 +\b c_2 =0$. Similarly we also have $-\b c_1 +\a c_2 =0$.
From both equations $\a c_1 +\b c_2 =0, \ -\b c_1 +\a c_2 =0$, it follows that
$c_1 =c_2 =0$, a contradiction.

As the code \C\ is of dimension $n+2$ and ${\mbox{dim}} \ \C + {\mbox{dim}}\ \C^\perp = 2n+4$,
$\C$ and $\C^\perp$ have the same dimension. Since $\C \subseteq \C^\perp,$
we have $\C = \C^\perp$, that is, \C\ is self-dual.
\end{proof}

We give a more efficient algorithm to construct $G$ in Proposition~\ref{prop:BU} as follows. The idea of this construction comes from the recursive algorithm in~\cite{AG},~\cite{AguGabKimSokSol}.

\medskip

{\bf{Modified building-up construction}}

\begin{itemize}

\item
Step 1:

Under the same notations as above, we consider the following.

For each $i$, let $s_i$ and $t_i$ be in $GF(q)$ and define
 $\by_i := ( s_i, t_i,
\a s_i+\b t_i, \b s_i-\a t_i)$ be a vector of length $4$.
Then
\[ G1 = \left[ \begin{array}{cccc|ccccccc}
 & \by_1 && & &  &&\br_1 &&& \\
 & \vdots && & &  && \vdots &&& \\
  & \by_n && & &  &&\br_n &&& \\
\end{array}
\right]
\]
generates a self-orthogonal code $C_1$.

\item Step 2:

Let $C$ be the dual of $C_1$.
Consider the quotient space $C/C_1$.
 Let $U_1$ be the set of all coset representatives of the form $\bx_1'=(1~0~0~0~ \bx_1)$ such that $\bx_1' \cdot \bx_1' =0$
 and  $U_2$ the set of all coset representatives of the form $\bx_2'=(0~1~0~0~ \bx_2)$ such that $\bx_2' \cdot \bx_2' =0$.

\item Step 3:

For any $\bx_1' \in U_1$ and $\bx_2' \in U_2$ such that $\bx_1' \cdot \bx_2'=0$,
the following matrix

\[ G = \left[ \begin{array}{cccc|ccccccc}
 1 &0 &0 &0 & & &&\bx_1 &&& \\
 0 &1 &0 &0 & & &&\bx_2 &&& \\ \hline
 & \by_1 && & &  &&\br_1 &&& \\
 & \vdots && & &  && \vdots &&& \\
  & \by_n && & &  &&\br_n &&& \\
\end{array}
\right]
\]

generates a self-dual code \C\ over $GF(q)$ of length $2n+4$.

\end{itemize}

Then, we have the following immediately.

\begin{proposition}
Let $SD_1$ be the set of all self-dual codes obtained from Proposition~\ref{prop:BU} with all possible vectors of $\bx_1$ and $\bx_2$. Let $SD_2$ be the set of all self-dual codes obtained from
the modified building-up construction with all possible values of $s_i$ and $t_i$ in $GF(q)$ for $1 \le i \le n$. Then $SD_1 = SD_2$.
\end{proposition}

What follows is the converse of Proposition~\ref{prop:BU}, that is,
every self-dual code over $GF(q)$ with $q \equiv 3 \pmod 4$ can be
obtained by the building-up method in Proposition~\ref{prop:BU}.

\begin{proposition}
\label{prop:converse} Let $q$ be a power of an odd prime such that
$q \equiv 3 \pmod 4$. Any self-dual code \C\ over $GF(q)$
of length $2n$ with even $n \ge 4$ is obtained
from some self-dual code $\C_0$ over $GF(q)$ of length
$2n-4$ (up to permutation equivalence) by the construction method given
in Proposition~\ref{prop:BU}.
\end{proposition}
\begin{proof}
Let $G$ be a generator matrix of \C. Without loss of generality we
may assume that $G=(I_n \mid A)=(\be_i \mid \ba_i)$, where $\be_i$
and $\ba_i$ are the row vectors of $I_n$ (= the identity matrix) and
$A$, respectively for $1 \le i \le n$. It is enough to show that
there exist vectors $\bx_1, \bx_2$ in $GF(q)^{2n-4}$ and a self-dual
code $\C_0$ over $GF(q)$ of length $2n-4$ whose extended code $\C_1$
(constructed by the method in Proposition~\ref{prop:BU}) is equivalent
to \C.

We note that $\ba_i \cdot \ba_j = 0$ for $i \not = j$, $1 \le i,j
\le n$ and $\ba_i \cdot \ba_i = -1$ for $1 \le i \le n$ since \C\ is
self-dual. Let $\alpha$ and $\beta$ be in $GF(q)^\ast$ such that
$\alpha^2+\beta^2+1=0$ in $GF(q)$. We notice that \C\ also has the
following generator matrix

\[ G^\prime := \left[ \begin{array}{c|c}
 \be_1 +\a\be_3 +\b\be_4 & \ba_1 +\a\ba_3 +\b\ba_4 \\
 \be_2 +\b\be_3 -\a\be_4 & \ba_2 +\b\ba_3 -\a\ba_4 \\
 \be_3  &  \ba_3\\ \be_4  & \ba_4 \\
 \vdots  & \vdots \\ \be_n  & \ba_n\\
\end{array}
\right].
\]

Deleting the first four columns and the third and fourth rows of
$G^\prime$ produces the following $(n-2)\times(2n-4)$ matrix $G_0$:

\[ G_0 := \left[ \begin{array}{ccc|c}
 0 & \cdots & 0 & \ba_1 +\a\ba_3 +\b\ba_4\\
 0 & \cdots & 0 & \ba_2 +\b\ba_3 -\a\ba_4 \\
 & & & \ba_5 \\ & I_{n-4} & & \vdots \\ & & & \ba_n\\
\end{array}
\right]
\]

We claim that $G_0$ is a generator matrix of some self-dual
code $\C_0$ of length $2n-4$. We first show that $G_0$ generates
a self-orthogonal code $\C_0$ as follows.
The inner product of the first row of $G_0$ with itself is equal to
$$\ba_1 \cdot\ba_1 + \a^2\ba_3\cdot\ba_3 + \b^2\ba_4\cdot\ba_4 =-(1+\a^2+\b^2)=0,$$
and similarly the second row is orthogonal to itself. For $3 \le i \le n-2$,
the inner product of the $i$th row of $G_0$ with itself equals
$1+ \ba_{i+2} \cdot \ba_{i+2}=0.$ The inner product of the first row of
$G_0$ with the second row is $\a\b\ba_3\cdot\ba_3-\a\b\ba_4\cdot\ba_4 =0.$
Clearly, for $1 \le i, j \le n-2$ with $i \neq j$,
any $i$th row is orthogonal to any $j$th row.

Now we show that $|\C_0|=q^{n-2}$, so $\C_0$ is self-dual. First of all, we note that
both vectors $\bv_1:=\ba_1 +\a\ba_3 +\b\ba_4$ and $\bv_2:= \ba_2 +\b\ba_3 -\a\ba_4$
in the first two rows of $G_0$ contain units. Otherwise, both vectors are zero vectors.
Then  $\ba_1 = - (\a\ba_3 +\b\ba_4)$, then $-1=\ba_1\cdot \ba_1 = (\a\ba_3 +\b\ba_4)\cdot (\a\ba_3 +\b\ba_4) =-(\a^2+\b^2) =1,$ i.e., $-1=1$ in $GF(q)$, which is impossible since $q$ is odd. So,
$\bv_1$ is a nonzero vector, and hence it contains a unit. Similarly, it is also true for $\bv_2$.
We can also show that $\bv_1$ and $\bv_2$ are linearly independent.
If not, $\bv_1= c \bv_2$ for some $c$ in $GF(q)^\ast$.
Then by taking inner products of both sides with $\ba_1$,
we have $\ba_1 \cdot \bv_1  = c\ba_1 \cdot \bv_2,$ so we get $-1 = 0,$
a contradiction. Therefore it follows that $G_0$ is equivalent to a standard form of matrix
$[I_{n-2} \mid \ast \ ]$, so that $|\C_0|=q^{n-2}$, that is, $\C_0$ is self-dual.

Let $\bx_1 =(0,\cdots,0 \mid \ba_1)$ and $\bx_2 =(0,\cdots,0 \mid
\ba_2)$ be row vectors of length $2n-4$. Then for $i=1, 2$,
$\mbox{\bf x}_i\cdot \mbox{\bf x}_i = \ba_i \cdot \ba_i = -1 \mbox{ in GF}(q)$
and $\mbox{\bf x}_1\cdot \mbox{\bf x}_2 = \ba_1 \cdot \ba_2 =0 \mbox{ in GF}(q)$.
Using the vectors $\bx_1, \bx_2$ and the self-dual code $\C_0$, we can
construct a self-dual code $\C_1$ with the following $n\times 2n$ generator
matrix $G_1$ by Proposition~\ref{prop:BU}:

\[ G_1 := \left[
\begin{array}{cccc|cccc}
 1 & 0 & 0 & 0 & 0 & \cdots & 0 & \ba_1\\
 0 & 1 & 0 & 0 & 0 & \cdots & 0 & \ba_2\\ \hline
 1 & 0 & \a & \b & 0 & \cdots & 0 & \ba_1 +\a\ba_3 +\b\ba_4 \\
 0 & 1 & \b & -\a & 0 & \cdots & 0 & \ba_2 +\b\ba_3 -\a\ba_4\\
 0 & 0  & 0 & 0  &  &   & & \ba_5\\
  \vdots & \vdots & \vdots & \vdots  & & I_{n-4} &  & \vdots \\
 0 & 0  & 0 & 0  &  &  & & \ba_n\\
\end{array}
\right]
\]

Clearly $G_1$ is row equivalent to $G$.
Hence the given code \C\ is the same as the code $\C_1$ that is
obtained from the code $\C_0$ by the building-up construction in
Proposition~\ref{prop:BU}. This completes the proof.
\end{proof}

\begin{remark}{\em
Note that in the statement of Proposition~\ref{prop:converse}
we do not have any condition on the minimum distance of $C$.
In the middle part of the proof of Proposition~\ref{prop:converse}
we have shown that $G_0$ has size $(n-2) \times (2n-4)$ and has
dimension $n-2$ without using the minimum distance of $C$.
}
\end{remark}
\subsection{Self-dual codes over $GF(3)$}
\label{subsec:eg_GF3} We consider self-dual codes over $GF(3)$. The
classification of extremal self-dual codes over $GF(3)$ was known up
to length $24$. For length $n=28$, only $32$ ternary extremal
self-dual codes were known~\cite{Huf_92},~\cite{Har_98} (or
see~\cite{Huf_05}). In particular, W. C. Huffman~\cite{Huf_92}
classified all $[28,14,9]$ self-dual ternary codes with a monomial
automorphism of prime order $\ge 5$ and showed that there are
exactly $19$ such codes. Using Proposition~\ref{prop:BU} with the
Pless symmetry code $\mathcal S(11)$ of length $24$
(see~\cite{Ple72},~\cite{HufPle}), we find easily at least $673$
inequivalent $[28,14,9]$ self-dual ternary codes whose full
automorphism group order is $2^{i+1}$, $i=0,1,2$. Note that any
ternary code has a trivial automorphism of order $2$. We list only
$20$ of them in order to save space in Table~\ref{tab:F3_28}, where
${\bf{x}_1}=(0 0 0 0 0 0 0 0 0 0 2 1 2 1 2 1 2 1 0 0 0 0 0 0)$ and
the $14$ entries of the right side of ${\bf{x}_2}$ are displayed in
the second column, and the order of the automorphism group of the
corresponding code is given in the last column. We note that by
Construction $A$ (see~\cite{ConSlo99},~\cite{HarKha}, or
Section~\ref{subsec:consA}, for example) the corresponding lattice
$\Lambda(C)$ of any ternary self-dual $[28,14,9]$ code $C$ produces
an optimal Type I $28$-dimensional unimodular lattice with minimum
norm $3$. On the other hand, Harada, Munemasa and Venkov have
recently verified that there are exactly $6,931$ extremal self-dual
codes over $GF(3)$~\cite{HarMunVen}, using the classification of all
the $28$-dimensional unimodular lattices with minimum norm $3$.

\begin{table}
\caption{Ternary $[28,14,9]$ self-dual codes using $\mathcal S(11)$
with {\small ${\bf{x}_1}=(0 0 0 0 0 0 0 0 0 0 2 1 2 1 2 1 2 1 0 0 0
0 0 0)$}} \label{tab:F3_28} \centering{\vspace{15pt}
\begin{tabular}{|c|c|c|}
\hline Code No. &
${\bf{x}_2}=(0 \dots 0 x_{11} \dots x_{24})$ & $|{\mbox{Aut}}|$\\
\hline
1 &  0 1 2 2 2 2 1 0 2 1 0 0 0 0 & 2\\
2 &  1 0 2 1 1 1 1 0 2 1 0 0 0 0& 2\\
3 &  2 2 1 0 1 1 2 0 2 1 0 0 0 0 & 2\\
4 &   1 2 1 0 2 1 2 0 2 1 0 0 0 0&2\\
5 &2 0 0 2 2 2 1 1 2 1 0 0 0 0 &2\\
6 &2 0 1 1 1 0 1 1 2 1 0 0 0 0 & 2\\
7& 2 0 2 1 1 0 1 2 2 1 0 0 0 0& 2\\
8& 2 0 1 1 2 0 1 2 2 1 0 0 0 0& 4\\
9& 0 1 1 1 2 0 1 2 2 1 0 0 0 0& 2\\
10&1 0 2 1 2 0 1 2 2 1 0 0 0 0& 2\\
11& 1 0 0 1 1 1 2 2 2 1 0 0 0 0&4\\
12&1 2 0 0 2 1 2 2 2 1 0 0 0 0&2 \\
13&0 1 1 2 2 2 0 1 1 1 0 0 0 0& 2\\
14&1 0 2 2 2 2 0 1 1 1 0 0 0 0&2\\
15&0 1 2 2 2 0 1 2 1 1 0 0 0 0& 2\\
16& 1 2 1 0 2 1 2 0 1 1 0 0 0 0&2\\
17&2 1 1 0 2 2 1 0 0 2 1 0 0 0&2\\
18&2 2 2 0 2 2 1 0 0 2 1 0 0 0&4\\
19&1 1 2 0 2 2 1 0 0 2 1 0 0 0&4\\
20&0 0 1 0 2 2 1 2 2 2 0 1 0 0&8\\
 \hline
\end{tabular}}
\end{table}



For length $n=32$, Huffman~\cite{Huf_92} classified all ternary
$[32,16,9]$ self-dual codes with a monomial automorphism of prime
order $r \ge 5$. He showed that $r$ can be assumed to be $r=5$ or
$r=7$. More precisely, he showed that there are exactly $239$
inequivalent extremal self-dual ternary $[32,16,9]$ codes with
monomial automorphisms of prime order $5$ and exactly $16$
inequivalent extremal self-dual ternary $[32,16,9]$ codes with
monomial automorphisms of prime order $7$. The equivalence between
these two classes of codes was not done. Only one extremal self-dual
$[32, 16, 9]$ code with a trivial automorphism group was found
in~\cite{Har_01}, but we have found a lot as shown below. Recently
Harada et.~al.~\cite{HarHozKhaKho} have found $53$ more inequivalent
extremal self-dual $[32, 16, 9]$ codes whose automorphism group
orders are divisible by $32$. Therefore the currently known number
of inequivalent extremal self-dual ternary $[32,16,9]$ codes is
$293$~\cite{HarHozKhaKho}.

Using Proposition~\ref{prop:BU} with a ternary self-dual $[28,14,9]$
code $C_{28}$ whose generator matrix $G(C_{28})$ is given below,
we find at least $945$ inequivalent $[32,16,9]$ self-dual ternary
codes, each of which has a trivial automorphism group.
These are not equivalent to the self-dual $[32,16,9]$ code with
a trivial automorphism group in~\cite{Har_01}. We have stopped
running Magma~\cite{Mag} and expect that there will be more such codes.
We list only $20$ of them in order to save space in Table~\ref{tab:F3},
where the $16$ entries of the right side of ${\bf{x}_2}$ are displayed in the
second column.

We summarize our result as follows.
\begin{proposition} There are at least $1238$ inequivalent extremal
ternary self-dual $[32, 16, 9]$ codes, $946$ of which have trivial
automorphism groups.
\end{proposition}

\begin{table}
\caption{New ternary $[32,16,9]$ self-dual codes with trivial
automorphism groups using $G(C_{28})$ with ${\bf{x}_1}=(0 0 0 0 0 0 0 0 0 0 0 0 2
1 2 1 2 1 2 1 0 0 0 0 0 0 0 0)$} \label{tab:F3}
\centering{\vspace{15pt}
\begin{tabular}{|c|c|}
\hline Code No. & ${\bf{x}_2}=(0 \dots 0 x_{13} \dots x_{28})$
 \\
\hline
1 & 1 2 1 1 2 1 0 0 2 1 0 0 0 0 0 0  \\
2 & 1 1 1 2 2 1 0 0 2 1 0 0 0 0 0 0  \\
3 & 2 2 1 2 2 1 0 0 2 1 0 0 0 0 0 0  \\
4 & 2 2 1 1 1 1 0 0 2 1 0 0 0 0 0 0  \\
5 & 1 1 1 1 1 1 0 0 2 1 0 0 0 0 0 0  \\
6 & 1 2 2 1 1 1 0 0 2 1 0 0 0 0 0 0  \\
7 & 2 2 2 2 1 1 0 0 2 1 0 0 0 0 0 0  \\
8 & 1 1 2 1 1 2 0 0 2 1 0 0 0 0 0 0  \\
9 & 2 2 2 1 1 2 0 0 2 1 0 0 0 0 0 0  \\
10& 1 1 2 2 2 2 0 0 2 1 0 0 0 0 0 0  \\
11& 2 2 2 2 2 2 0 0 2 1 0 0 0 0 0 0  \\
12& 1 1 1 1 2 2 0 0 2 1 0 0 0 0 0 0  \\
13& 2 2 1 1 2 2 0 0 2 1 0 0 0 0 0 0  \\
14& 1 2 2 1 2 2 0 0 2 1 0 0 0 0 0 0  \\
15& 1 1 2 1 0 2 1 0 2 1 0 0 0 0 0 0  \\
16& 2 2 2 1 0 2 1 0 2 1 0 0 0 0 0 0  \\
17& 0 1 2 1 1 2 1 0 2 1 0 0 0 0 0 0  \\
18& 1 1 0 1 2 2 1 0 2 1 0 0 0 0 0 0  \\
19& 0 1 1 1 2 2 1 0 2 1 0 0 0 0 0 0  \\
20& 0 1 2 2 2 2 1 0 2 1 0 0 0 0 0 0  \\
 \hline
\end{tabular}}
\end{table}

\[G(\C_{28})= \left[ \begin{array}{c}
1 0 0 0 0 0 0 0 0 0 0 0 0 0 2 1 2 1 2 1 2 1 0
0 0 0 0 0 \\
0 1 0 0 0 0 0 0 0 0 0 0 0 0 0 1 2 2 2 2 1 0 2
1 0 0 0 0 \\
2 2 1 0 1 0 0 0 0 0 0 0 0 0 0 0 0 1 1 1 1 1 1
1 1 1 1 1 \\
1 0 1 1 0 1 0 0 0 0 0 0 0 0 0 0 2 0 1 2 1 1 1
2 2 2 1 2 \\
0 0 0 0 0 0 1 0 0 0 0 0 0 0 0 0 2 2 0 1 2 1 1
1 2 2 2 1 \\
2 1 0 1 0 0 0 1 0 0 0 0 0 0 0 0 2 1 2 0 1 2 1
1 1 2 2 2 \\
1 2 0 2 0 0 0 0 1 0 0 0 0 0 0 0 2 2 1 2 0 1 2
1 1 1 2 2 \\
0 1 1 2 0 0 0 0 0 1 0 0 0 0 0 0 2 2 2 1 2 0 1
2 1 1 1 2 \\
2 0 2 2 0 0 0 0 0 0 1 0 0 0 0 0 2 2 2 2 1 2 0
1 2 1 1 1 \\
2 1 0 1 0 0 0 0 0 0 0 1 0 0 0 0 2 1 2 2 2 1 2
0 1 2 1 1 \\
0 0 0 0 0 0 0 0 0 0 0 0 1 0 0 0 2 1 1 2 2 2 1
2 0 1 2 1 \\
1 1 2 0 0 0 0 0 0 0 0 0 0 1 0 0 2 1 1 1 2 2 2
1 2 0 1 2 \\
0 2 2 1 0 0 0 0 0 0 0 0 0 0 1 0 2 2 1 1 1 2 2
2 1 2 0 1 \\
1 1 2 0 0 0 0 0 0 0 0 0 0 0 0 1 2 1 2 1 1 1 2
2 2 1 2 0 \\
\end{array} \right]
\]

\subsection{Self-dual codes over $GF(7)$}
\label{subsec:eg_GF7}

Next we consider self-dual codes over $GF(7)$. The classification of
self-dual codes over $GF(7)$ was known up to lengths $12$
(see~\cite{GulHar99, GulHarMiy, HarOst, PleTon}).
The papers~\cite{GulHar99, GulHarMiy} used the monomial equivalence and monomial automorphism groups of self-dual codes over $GF(7)$. Hence we also use the monomial equivalence and monomial automorphism groups.
On the other hand, the $(1,-1,0)$-monomial equivalence was used in~\cite[Theorem 1]{PleTon} to give a mass formula:

\[
\sum_{\substack{j}} \frac{2^n n!}{|Aut(C_j)|} =N(n)=  2 \prod_{i=1}^{(n-2)/2} (7^i+1),
\]
where $N(n)$ denotes the total number of distinct self-dual codes over $GF(7)$. In particular, when $n=16$, there are at least $785086 >N(16)/{2^{16} 16!}$ inequivalent self-dual $[16, 8]$ codes over $GF(7)$ under the $(1,-1,0)$-monomial equivalence. It will be very difficult to classify all self-dual $[16, 8]$ codes. In what follows, we focus on self-dual codes with the highest minimum distance.

\medskip

For length $n=16$,
only ten optimal self-dual $[16,8,7]$ codes over $GF(7)$ were
known~\cite{GulHarMiy}. These have (monomial) automorphism group orders $96$ or
$192$. We construct at least $214$ self-dual $[16,8,7]$ codes over
$GF(7)$ by applying the building-up construction to the bordered
circulant code with $\alpha=0, \beta=2=\gamma$ and the row
$(2,5,5,2,0)$, denoted by $C_{1,1}$ in~\cite{GulHar99}. We check
that the $207$ codes of the $214$ codes have automorphism group
orders $6,12,24,48,72$, and hence they are new. On the other hand,
the remaining seven codes have group orders $96$ or $192$, and we
have checked that six of them are equivalent to the first four codes
and the last two codes in~\cite[Table 7]{GulHarMiy}, and that the
remaining one code is new. We list $20$ of our $214$ codes in
Table~\ref{tab:F7_16}, where ${\bf{x}}_1$ and ${\bf{x}}_2$ are given
in the second and third columns respectively, and $A_7$ and $A_8$
are given in the last column so that the Hamming weight enumerator
of the corresponding code can be derived from the appendix
of~\cite{GulHarMiy}.

\begin{theorem}
There exist at least $218$ self-dual $[16,8,7]$ codes over $GF(7)$.
\end{theorem}

For length $20$ only one optimal self-dual $[20,10,9]$ code over $GF(7)$ is
known~\cite{GulHar99},~\cite{GulHarMiy}. It is an open question to determine
whether this code is unique.

For length $24$ there are $488$ best known self-dual $[24, 12,9]$
codes over $GF(7)$~\cite{GulHarMiy}. It has been
confirmed~\cite{Har_09} that the $488$ codes in~\cite{GulHarMiy}
(only $40$ codes are shown in~\cite{GulHarMiy}) have non-trivial
automorphism groups. On the other hand, we have found at least $59$
self-dual $[24, 12, 9]$ codes over $GF(7)$, each of which has a
trivial automorphism group. To do this, we have used the bordered
circulant code over $GF(7)$ with $\alpha=2, \beta=1=\gamma$ and the
row $(4,6,3,6,6,1,4,3,0)$, denoted by $C_{20,1}$~\cite{GulHar99}. We
list $10$ of our $59$ codes in Table~\ref{tab:F7_20}, where
${\bf{x}}_1$ and ${\bf{x}}_2$ are given in the second and third
columns respectively, and $A_9, \dots, A_{12}$ are given in the last
column so that the Hamming weight enumerator of the corresponding
code can be derived from the appendix of~\cite{GulHarMiy}. We
therefore obtain the following theorem.

\begin{theorem}
There exist at least $547$ self-dual $[24,12,9]$ codes over $GF(7)$.
\end{theorem}

\begin{table}
\caption{New $[16,8,7]$ self-dual codes over $GF(7)$ using $C_{1,1}$
in~\cite{GulHar99}} \label{tab:F7_16} \centering{\vspace{15pt}
\begin{tabular}{|c|c|c|c|c|}
\hline $\#$ & ${\bf{x}_1}=(0 \dots 0 x_1 \dots x_{12})$
&
${\bf{x}_2}=(0 \dots 0 x_{5} \dots x_{12})$ & $|{\mbox{Aut}}|$ & $A_7, A_8$\\
\hline
1 &  2 1 2 6 1 6 1 0 &  1 2 1 1 6 5 1 0 & 24 & 696, 3432\\
2 &  1 2 2 6 1 6 1 0 &  4 5 6 4 4 6 1 0 &
24 &720, 3360 \\
3 &  5 1 5 6 1 6 1 0 &  4 5 1 3 6 1 3 0& 12 &  636, 3780\\
4& 5 1 5 1 1 6 1 0& 6 3 3 6 1 2 3 0 &6 &  564, 3996\\
5& 6 5 5 1 1 6 1 0& 3 4 1 2 4 1 1 0&12 & 540, 4068\\
6& 5 2 1 1 1 6 1 0&  2 1 2 1 5 2 3 0&12& 588, 3924\\
7& 1 6 2 2 1 6 1 0&  3 2 1 5 1 2 2 0&6 & 612, 3804\\
8& 4 2 3 3 1 6 1 0&  3 3 5 3 3 5 2 0&12& 576, 3936\\
9& 5 3 3 3 1 6 1 0&  4 1 4 5 1 3 1 0&12& 588, 3876\\
10& 3 2 4 3 1 6 1 0& 5 5 2 4 1 5 1 0&12& 552, 4104\\
11& 2 3 4 3 1 6 1 0& 4 4 5 4 4 2 2 0&12& 624, 3744\\
12& 5 4 4 3 1 6 1 0& 3 6 2 6 3 1 3 0&12& 612, 3852\\
13& 5 3 4 4 1 6 1 0& 5 5 5 3 5 1 1 0&48& 576, 3936\\
14& 1 5 1 5 1 6 1 0& 3 1 1 2 4 3 1 0&24& 480, 4320\\
15& 2 6 1 5 1 6 1 0& 5 3 1 1 1 3 3 0&24& 672, 3552\\
16& 3 4 4 5 1 6 1 0& 5 2 5 3 6 2 1 0&48& 528, 4128\\
17& 2 1 6 5 1 6 1 0& 6 2 5 2 3 2 1 0&12& 672, 3552\\
18& 5 2 3 5 2 6 1 0& 1 4 4 5 1 4 1 0&12& 660, 3708\\
19& 2 2 4 5 2 6 1 0& 2 1 2 1 2 5 3 0&6 & 564, 4092\\
20& 6 6 6 5 2 6 1 0& 1 3 1 4 6 2 3 0&6 & 600, 3912\\
 \hline
\end{tabular}}
\end{table}

\begin{table}
\caption{New $[24,12,9]$ self-dual codes over $GF(7)$ using
$C_{20,1}$} in~\cite{GulHar99} with trivial automorphism groups
\label{tab:F7_20} \centering{\vspace{15pt}
\begin{tabular}{|c|c|c|c|}
\hline $\#$ & ${\bf{x}_1}=(0\dots 0 x_9 \dots x_{20})$
&
${\bf{x}_2}=(0 \dots 0 x_{9} \dots x_{20})$ & $A_9, A_{10}, A_{11}, A_{12}$\\
\hline
1 & 2 6 2 3 2 1 6 1 6 1 0 0 &
    4 4 3 5 3 2 1 1 6 1 0 0 &  948, 8496, 65520, 425484 \\
2 &2 2 5 1 3 1 6 1 6 1 0 0 &
 3 5 4 4 6 4 2 1 6 1 0 0 &  894, 8802,  64572, 427236 \\
3 & 6 4 4 1 4 1 6 1 6 1 0 0 &
3 6 2 6 1 2 2 1 6 1 0 0 &  936, 8436, 65580, 427704 \\
4& 2 6 2 3 5 1 6 1 6 1 0 0 &
   5 3 3 4 4 2 1 1 6 1 0 0 &  882, 8592, 65544, 427086 \\
5& 5 6 5 4 5 1 6 1 6 1 0 0 &
   2 1 3 5 1 5 1 1 6 1 0 0& 774, 8706, 66204, 426204 \\
6 &1 4 2 2 1 2 6 1 6 1 0 0&
 3 3 5 6 3 4 2 1 6 1 0 0 &  948, 8466, 65520, 426306 \\
7 &4 5 3 4 4 2 6 1 6 1 0 0 &
 1 3 5 1 2 1 2 1 6 1 0 0& 936, 8982, 63516, 426750 \\
8 &1 6 4 6 4 3 6 1 6 1 0 0&
 2 1 6 3 2 6 2 1 6 1 0 0&  966, 8502, 65148, 426792 \\
9 &1 3 3 1 1 3 6 1 6 1 0 0&
 5 2 2 3 2 4 2 1 6 1 0 0& 966, 8700, 64500, 425730 \\
10 &4 6 1 6 3 4 6 1 6 1 0 0&
 5 1 6 3 6 2 2 1 6 1 0 0& 846, 8796, 65448, 424134 \\
 \hline
\end{tabular}}
\end{table}

\section{Building-up construction for self-dual
codes over finite chain rings} \label{sec:building_rings}

\subsection{Finite chain rings}\label{sec:chainring}

A finite commutative ring with identity $\ne 0$ is called a \textit{chain ring} if its ideals are
linearly ordered by inclusion. This means that it has a unique
maximal ideal, i.e., that it is a local ring.
Let $R$ be a finite chain ring, ${\mathfrak m}$ the unique maximal
ideal of $R$, and $\gamma$ the generator of the unique
maximal ideal ${\mathfrak m}$. Then ${\mathfrak
m}=\langle\gamma\rangle=R\gamma$, where
$R\gamma=\langle\gamma\rangle=\{\beta\gamma\,|\,\beta\in R\}$. We
have
$R=\langle\gamma^0\rangle\supseteq\langle\gamma^1\rangle\supseteq\cdots
\supseteq\langle\gamma^i\rangle\supseteq\cdots.$
This chain cannot be infinite, since $R$ is
finite. Therefore, there exists a positive integer $i$ such that
$\langle\gamma^i\rangle=\{0\}$. Let $e$ be the minimal number such
that $\langle\gamma^e\rangle=\{0\}$. We call $e$ the \textit{nilpotency
index} of $\gamma$.

Let $C$ be a linear code over a finite chain ring $R$ of length $n$. Then its generator matrix is equivalent to the following generator matrix $G$:
\begin{equation*}\label{pm-1}
G=\left[
\begin{array}{cccccc}I_{k_0}&A_{0,\,1}&A_{0,\,2}&\cdots&A_{0,\,e-1}&A_{0,\,e}\cr
0&\gamma I_{k_1}&\gamma A_{1,\,2}&\cdots&\gamma A_{1,\,e-1}&\gamma
A_{1,\,e}\cr 0&0&\gamma^2
I_{k_2}&\cdots&\gamma^2A_{2,\,e-1}&\gamma^2A_{2,\,e}\cr
\cdots&\cdots&\cdots&\cdots&\cdots&\cdots \cr
0&0&\cdots&\cdots&\gamma^{e-1}I_{k_{e-1}}&\gamma^{e-1}A_{e-1,\,e}
\end{array} \right].
\end{equation*}

Let $|R|$ denote the cardinality of $R$ and $R^{\ast}$  the set
of all units in $R$. We know that $R^{\ast}$ is a multiplicative
group under the multiplicative operation of $R$. Let
$\mathbb{F}=R/{\mathfrak m}=R/\langle\gamma\rangle$ be the residue field
with characteristic $p$, where $p$ is a prime number. This implies
that  there exist integers $q$ and $r$ such that $|\mathbb{F}|=q=p^r$,
and $\mathbb{F}^{\ast}=\mathbb{F}-\{0\}$. This implies that
$|\mathbb{F}^{\ast}|=p^r-1$. See~\cite{Ana} for codes over chain rings.

The following theorem is the building-up construction for self-dual
codes over a finite chain ring $R$ with the property that there
exist $\alpha$ and $\beta$ in $R^\ast$ such that
$\alpha^2+\beta^2+1=0$ in $R$.

\begin{proposition}
\label{prop:BU_2} Let $R$ be a finite chain ring. Suppose that there
exist $\alpha$ and $\beta$ in $R^\ast$ such that
$\alpha^2+\beta^2+1=0$ in $R$. Let $G_0=(\br_i)$ be a generator
matrix (not necessarily in standard form) of a self-dual code $\C_0$
over $R$ of length $2n$, where $\br_i$ are the row vectors with $1
\le i \le k$ for some positive integer $k$. Let $\mbox{\bf x}_1$ and
$\mbox{\bf x}_2$ be vectors in $R^{2n}$ such that $\mbox{\bf
x}_1\cdot\mbox{\bf x}_2 = 0$ in $R$ and $\mbox{\bf
x}_i\cdot\mbox{\bf x}_i = -1$ in $R$ for each $i=1,2$. For each $i$,
$ 1 \le i \le k$, let $s_i := \mbox{\bf x}_1 \cdot \br_i$, $t_i:=
\mbox{\bf x}_2 \cdot \br_i$, and $\by_i := ( -s_i, -t_i, -\a s_i-\b
t_i, -\b s_i+\a t_i)$ be a vector of length $4$. Then the following
matrix

\[ G = \left[ \begin{array}{cccc|ccccccc}
 1 &0 &0 &0 & & &&\bx_1 &&& \\
 0 &1 &0 &0 & & &&\bx_2 &&& \\ \hline
 & \by_1 && & &  &&\br_1 &&& \\
 & \vdots && & &  && \vdots &&& \\
  & \by_k && & &  &&\br_k &&& \\
\end{array}
\right]
\]

generates a self-dual code \C\ over $R$ of length $2n+4$.
\end{proposition}
\begin{proof}
The proof is very similar to that of Proposition~\ref{prop:BU}.
It is straightforward to see that $\C$ is self-orthogonal, so $\C \subseteq \C^\perp.$
By the exactly same reasoning as the proof of Proposition~\ref{prop:BU},
we can show that no linear combination of the first two rows of $G$
(with scalars in $R$) is in the span of the bottom $n$ rows of $G$.
It thus follows that $|\C| = |R|^2 |\C_0|$. Since $|\C_0| = |R|^n$,
we have $|\C| = |R|^{n+2}$. Furthermore,
we have $|\C||\C^\perp| = |R|^{2n+4}$, so $|\C| = |\C^\perp|$.
As $\C \subseteq \C^\perp$ and $|\C| = |\C^\perp|$, we have $\C = \C^\perp$,
that is, \C\ is self-dual.
\end{proof}

The following proposition shows that the converse of
Proposition~\ref{prop:BU_2} also holds for chain rings where there
exist $\alpha$ and $\beta$ in $R^\ast$ such that
$\alpha^2+\beta^2+1=0$ in $R$. That is, every self-dual code over
such a chain ring can be obtained by the method given in
Proposition~\ref{prop:BU_2}. In fact, the following result over
chain rings is a general version of Proposition~\ref{prop:converse}
over finite fields, and its proof requires the property of chain
rings. Proposition~\ref{prop:converse} is certainly a corollary of
Proposition~\ref{prop:converse_2}, but the proof of
Proposition~\ref{prop:converse} is simpler than that of
Proposition~\ref{prop:converse_2}; thus we treated the finite field
case in Section~\ref{sec:building-up} separately due to its
simplicity.

\begin{proposition}
\label{prop:converse_2} Let $R$ be a finite chain ring. Suppose that
there exist $\alpha$ and $\beta$ in $R^\ast$ such that
$\alpha^2+\beta^2+1=0$ in $R$. Any self-dual code \C\ over $R$ of
length $2n$ with $n$ even $ \ge 4$ and free rank $\ge 4$ is obtained from some self-dual code $\C_0$ over
$R$ of length $2n-4$ (up to permutation equivalence) by the
construction method given in Proposition~\ref{prop:BU_2}.
\end{proposition}
\begin{proof}
It is sufficient to show that there exist vectors $\bx_1, \bx_2$ in
$R^{2n-4}$ and a self-dual code $\C_0$ over $R$ of length $2n-4$
whose extended code $\C_1$ (constructed by the method in
Proposition~\ref{prop:BU_2}) is equivalent to \C. Let $G$ be a
generator matrix of \C\ in a standard form as follows:

\[ G := \left[
\begin{array}{ccccc}
 1 & 0 & 0 & 0 & \ba_1\\
 0 & 1 & 0 & 0 & \ba_2\\
 0 & 0 & 1 & 0 & \ba_3 \\
 0 & 0 & 0 & 1 & \ba_4\\ \hline
 0 & 0 & 0 & 0 & \ba_5\\
  \vdots & \vdots & \vdots &  \vdots \\
 0 & 0  & 0 & 0 & \ba_k
\end{array}
\right].
\]

Clearly \C\ also has the following generator matrix
$G^\prime$:

\[ G^\prime := \left[
\begin{array}{cccc|c}
 1 & 0 & 0 & 0  & \ba_1\\
 0 & 1 & 0 & 0  & \ba_2\\
 1 & 0 & \a & \b  & \ba_1 +\a\ba_3 +\b\ba_4 \\
 0 & 1 & \b & -\a &  \ba_2 +\b\ba_3 -\a\ba_4\\ \hline
 0 & 0  & 0 & 0  &  \ba_5\\
  \vdots & \vdots & \vdots & \vdots   & \vdots \\
 0 & 0  & 0 & 0  & \ba_k
\end{array}
\right].
\]

Deleting the first four columns and the first and second rows of
$G^\prime$ produces the following $(k-2)\times(2n-4)$ matrix $G_0$:

\[ G_0 := \left[ \begin{array}{c}
  \ba_1 +\a\ba_3 +\b\ba_4\\
 \ba_2 +\b\ba_3 -\a\ba_4 \\
 \ba_5 \\
 \vdots \\ \ba_k\\
\end{array}
\right].
\]

We claim that $G_0$ is a generator matrix of some self-dual code
$\C_0$ of length $2n-4$. First of all, we observe that $G_0$
generates a self-orthogonal code $\C_0$; this follows easily from
the following facts: $\ba_i\cdot\ba_j =0$ for $1\le i < j \le k,$
$\ba_i\cdot\ba_i =0$ for $5\le i \le k, \ba_i\cdot\ba_i = -1$ for
$1\le i \le 4,$ and $\a^2+\b^2+1=0$ in $R.$
Next we note that the $R$-span of the bottom $k-4$ rows of $G$ has
size $|R|^{n-4}$ as the first $4$ rows of $G$ have $R$-span size $|R|^4$.
Thus the $R$-span of the bottom $k-4$ rows of $G_0$ also has size $|R|^{n-4}$.
Hence to show that $|\C_0|=|R|^{n-2}$, we prove that
(a) both vectors $\bv_1:=\ba_1 +\a\ba_3 +\b\ba_4$ and $\bv_2:= \ba_2 +\b\ba_3
-\a\ba_4$ give free rank $2$ (that is, the $R$-span of $\{ \bv_1, \bv_2 \}$ has size $|R|^2$)
and that (b) only the zero vector in the $R$-span of $\{\bv_1, \bv_2 \}$ is in the $R$-span of
$\{\ba_5, \dots, \ba_k \}$.

For the part (a), unlike the finite field case, showing that
$\{\bv_1, \bv_2\}$ is linearly independent over $R$ is insufficient
since the $R$-span of $\{\bv_1, \bv_2\}$ does not necessarily give
size $|R|^2$. Instead we show in detail that $\bv_1$ and $\bv_2$
give free rank $2$ as follows. We first note that both vectors
$\bv_1, \bv_2$ contain unit components. If not, i.e., $\bv_1$
contains no unit components, then $\bv_1 = \gamma \bw$ for some
$\bw$ in $R^{2n-4}$ with $\gamma$ the generator of the unique
maximal ideal ${\mathfrak m}$ of $R$; so $\ba_1 \cdot \ba_3 =
(-\a\ba_3 - \b\ba_4 + \gamma \bw)\cdot \ba_3$. Thus we get $-\a =
\gamma (\bw \cdot \ba_3),$ and this shows that a unit $-\a$ is
contained in ${\mathfrak m}$, a contradiction. Similarly, it also
holds for $\bv_2$. In fact, both $\bv_1, \bv_2$ contain at least two
unit components; otherwise, $\bv_1$ has only one unit component, say
$u_1$. Then since $\bv_1 \cdot \bv_1 = 0$, we have $u_1^2 + \gamma z
=0$ for some $z$ in $R,$ which implies $u_1^2 \in {\mathfrak m}$, a
contradiction. This is also true for $\bv_2.$ Furthermore, we can
show that $\bv_1 \neq u \bv_2$ for any $u$ in $R^\ast$ in exactly
the same way as in Proposition~\ref{prop:converse}. Hence, it
follows that the $R$-span of $\{ \bv_1, \bv_2 \}$ is free of
rank $2$. For the part (b), suppose that
$c_1 \bv_1 + c_2 \bv_2= \sum_{i=5}^k b_i
\ba_i$ where $b_i \in R$ for $5 \le i \le k$. Then for $j=1, 2$,
$-c_j=(c_1 \bv_1 + c_2 \bv_2) \cdot \ba_j$ but $(\sum_{i=5}^k b_i
\ba_i) \cdot \ba_j =0$. Hence $c_j=0$ for $j=1,2$ as required.

Therefore we have $|\C_0|=|R|^{n-2}$, that is, $\C_0$ is self-dual.
The rest of the proof is the same as that of Proposition~\ref{prop:converse}.
\end{proof}

\subsection{Galois Rings}

One of the important examples of chain rings is a Galois ring.
In~\cite{KimLee2} we give the building-up method for self-dual codes
over Galois rings $\GR(p^m, r)$ in all the cases except the case $p
\equiv 3 \pmod{4}$ with $r$ odd. We complete the missing case by
using Proposition~\ref{prop:BU_2} and
Proposition~\ref{prop:converse_2} as follows.

\begin{proposition}
\label{prop:GaloisRing} The building-up method works over any Galois
ring $\GR(p^m, r)$ with $p$ an odd prime. More preciesly, if $p \equiv
1 \pmod{4}$, then the building-up method is given
by~\cite[Proposition 3.3, 3.4]{KimLee2}, and if $p \equiv 3
\pmod{4}$, then the building-up method is given by
Proposition~\ref{prop:BU_2},~\ref{prop:converse_2}.
\end{proposition}
\begin{proof}
It suffices to show it for the case $p \equiv 3 \pmod{4}$. By
Propositions~\ref{prop:BU_2} and~\ref{prop:converse_2}, we know
that the building-up method works over Galois rings $\GR(p^m, r)$
if there exist $\alpha$ and $\beta$ in $\GR(p^m, r)^\ast$ such
that $\alpha^2+\beta^2+1=0$ in $\GR(p^m, r)$. In fact, we have
$\Z_{p^m} \subseteq \GR(p^m, r)$. It is therefore enough to show
that when $p \equiv 3 \pmod{4}$, there exist $\alpha$ and
$\beta$ in $(\Z_{p^m})^\ast$ such that $\alpha^2+\beta^2+1=0$ in
$\Z_{p^m}$. If $p \equiv 3 \pmod{4}$, then by Lemma~\ref{lemma1}
there exist $\alpha$ and $\beta$ in $\Z_{p}^\ast$ such that
$\alpha^2+\beta^2+1=0$ in $\Z_{p}$.
We notice that $2$ and $\alpha$ are units in $\Z_{p^i}$
for any positive integer $i$. From~\cite[Lemma 3.9]{DouPar}, it
follows that $x_m^2+y_m^2+1 = 0$ in $\Z_{p^m}$ for any integer $m
\ge 1$, where $x_m$ and $y_m$ are defined recursively as follows:
We first let  \vspace{-0.2cm}
\begin{align}\notag
x_1 &= \alpha, \ \ y_1 = \beta, \ \ r_1= (x_1^2+y_1^2+1)/p, \\ \notag
\tilde{r}_1 &\equiv -\frac{r_1}{2\alpha} \pmod{p}, \ \ {\mbox{ where }} 0\le \tilde{r}_1 <p, \\ \notag
x_2 &= x_1+ \tilde{r}_1 p, \ \ \  y_2 =\b.
\end{align}
An easy calculation shows that $x_2^2+y_2^2+1 \equiv 0 \pmod{p^2}$ and
$x_2, y_2 \in \Z^\ast_{p^2}$. Assuming that there exist $x_{i-1}, y_{i-1} \in \Z^\ast_{p^{i-1}}$
such that $x_{i-1}^2+y_{i-1}^2+1 \equiv 0 \pmod{p^{i-1}}$ and $x_{i-1} \equiv \a \pmod{p},$
we recursively define
$r_{i-1} = (x_{i-1}^2+y_{i-1}^2+1) /{p^{i-1}}, \ \tilde{r}_{i-1} \equiv -\frac{r_{i-1}}{2\alpha} \pmod{p}$
where $0 \le \tilde{r}_i <p,$ $x_i = x_{i-1} + \tilde{r}_{i-1} p^{i-1},$ and $\ y_i = \beta.$
A straightforward calculation shows that $x_i \equiv \a \pmod{p}, \ x_i^2+y_i^2+1 \equiv 0 \pmod{p^i},$
and $x_i, y_i \in \Z^\ast_{p^i}$. In particular  \vspace{-0.2cm}
\begin{equation*}\label{eq_1}
x_m = \alpha + \tilde{r}_1 p + \tilde{r}_2 p^{2} + \cdots +
\tilde{r}_{m-1} p^{m-1}, \ \ \ y_m=\beta,
\end{equation*}
and we have $x_m^2+y_m^2+1 = 0$ in $\Z_{p^m}$.
\end{proof}


\noindent
\subsection{Self-dual codes over $\Z_9$ and their lattices}
\label{subsec:consA} In this section we consider self-dual codes
over a Galois ring $R=\GR(3^2, 1)= \Z_9$ and reconstruct optimal
Type I lattices of dimensions $12, 16, 20,$ and $24$ using
Construction $A$, which is described below
(see~\cite{BanDouHarOur99, ConSlo99, Gab04}).

\begin{definition}(Construction $A$) Let $m$ be any integer greater than $1$.
If $C$ is a self-dual code of length $n$ over $\mathbb Z_{m}$, then the lattice
\[\Lambda(C) = \frac{1}{\sqrt{m}}\{{\bf{x}}=(x_1, \dots, x_n) \in \mathbb Z^n~|~(x_1~({\mbox{mod}}~m),
\dots, x_n~({\mbox{mod}}~m)) \in C \}
\]
is an $n$-dimensional unimodular lattice with the minimum norm $\mu =\min\{\frac{d_E(C)}{m},~ m \}$,
where $d_E(C)$ denotes the minimum Euclidean weight of $C$.
\end{definition}

From Proposition~\ref{prop:GaloisRing} there exist $\alpha$ and
$\beta$ in $R^\ast$ such that $\alpha^2+\beta^2+1=0$ in $R$. We take
$\alpha =2$ and $\beta=2$. For example, $\{(1, 0, 2, 2),
(0,1,2,-2)\}$ generates a self-dual code $\C_1$ over $\Z_9$ of
length $4$ with minimum Hamming weight $3$.

By using Proposition~\ref{prop:BU_2} starting from $\C_1$ with
$\bx_1 = (1, 3, 5, 0)$ and $\bx_2 = (3, 8, 0, 4)$, we find the following
generator matrix $G_2$ of the self-dual code $\C_2$ over
$\Z_9$ of length $8$ with minimum Hamming weight $3$.

\[  G_2 = \left [
\begin{array}{cccccccccccccccccccc}
1 \ 0  \ 0 \  0  \ 1  \ 3  \ 5  \ 0\\
0 \  1  \ 0 \  0  \ 3  \ 8  \ 0  \ 4\\
7 \  7  \ 1 \  0  \ 1  \ 0  \ 2  \ 2\\
5 \  0  \ 1 \  1  \ 0  \ 1  \ 2  \ 7\\
\end{array}
\right].
\]

Its Hamming weight enumerator is $W_2(x,y)=x^8 + 16x^5y^3+
48x^4y^4 + 240x^3y^5 + 1072x^2y^6 + 2688xy^7+ 2496y^8$.

In what follows, we construct free self-dual codes over $\Z_9$ of lengths $12, 16,$ and $20$
all with minimum Hamming weight $6$. These codes can be regarded as codes over $GF(3)$ by
taking each coordinate modulo $3$. It is easy to see that the latter codes, called the {\it residue
codes} ${\mbox{Res}}(\C)$, are self-dual over $GF(3)$. In general, one can show that the residue code
${\mbox{Res}}(\C)$ of a {\it{free}} self-dual code $\C$ over $\Z_9$ is also self-dual over $GF(3)$
and that the minimum Hamming weight $d(\C)$ is the same as that of ${\mbox{Res}}(\C)$.
(In fact, it is known~\cite{DouKimLiu, Ana} that
$d(\C)=d({\mbox{Tor}}(\C))$ where ${\mbox{Tor}}(\C)=\{{\bf{v}} \pmod{3}~|~ 3 {\bf{v}} \in \C \}$.
Since ${\mbox{Tor}}(\C)={\mbox{Res}}(\C)$ for a free self-dual code $\C$ over $\Z_9$, the claim follows.)
Our self-dual codes over $\Z_9$ given below will attain
the highest possible minimum Hamming weight $6$ which free self-dual codes over $\Z_9$ of
lengths $12, 16,$ and $20$ can attain; it was known that the largest Hamming weight of
self-dual codes over $GF(3)$ of lengths $12, 16,$ and $20$ is $6$~\cite{Huf_05}.

Applying Proposition~\ref{prop:BU_2} to $G_2$, we obtain self-dual
codes of length $12$ with Hamming weight $6$. We list eight
inequivalent self-dual codes in Table~\ref{tab:Z9_1}, where the six
entries of the right side of ${\bf{x}_1}$ and ${\bf{x}_2}$
respectively are displayed in the second column and in the third column, the
fourth column gives the number $A_6$ of codewords with minimum
weight $6$, and the last column gives the minimum norm of the
corresponding lattice. By Construction $A$, we obtain the unique
optimal Type I lattice of dimension $12$~\cite{ConSlo99, Gab04}. As
far as we know, only one self-dual code over $\Z_9$ of length $12$
with Hamming weight $6$ is obtained by lifting the extended ternary
Golay $[12,6,6]$ linear code to a code over $\Z_9$~\cite{CalSlo96,
GreVit99}, and this code has $A_6=264$, which shows that our codes
in Table~\ref{tab:Z9_1} are certainly new.

In particular, the first code in Table~\ref{tab:Z9_1} has generator matrix given as follows:

\[  G_3 = \left [
\begin{array}{cccccccccccc}
1 \ 0 \ 0 \ 0 \ 0 \ 0 \ 4 \ 5 \ 1 \ 1 \ 1 \ 0 \\
0 \ 1 \ 0 \ 0 \ 0 \ 0 \ 2 \ 2 \ 2 \ 7 \ 0 \ 1 \\
0 \ 4 \ 8 \ 1 \ 1 \ 0 \ 0 \ 0 \ 1 \ 3 \ 5 \ 0 \\
7 \ 6 \ 8 \ 2 \ 0\ 1\ 0\ 0\ 3\ 8\ 0\ 4 \\
2\ 3\ 1\ 7\ 7\ 7\ 1\ 0\ 1\ 0\ 2\ 2 \\
6\ 0\ 3\ 3\ 5\ 0\ 1\ 1\ 0\ 1\ 2\ 7 \\
\end{array}
\right].
\]

Similarly, using Proposition~\ref{prop:BU_2} with $G_3$, we obtain
many inequivalent self-dual codes of length $16$ with Hamming weight
$6$ and $A_6=230+ 6t$ for $t=0, 1, \dots, 19$. Table~\ref{tab:Z9_2}
presents twenty of them, where the eight entries of the right side
of ${\bf{x}_1}$ and ${\bf{x}_2}$ respectively are displayed in the
second and the third column. By Construction $A$, we obtain the
unique optimal Type I lattice of dimension $16$~\cite{ConSlo99, Gab04}.
As an example, the self-dual code $\C_4$ (denoted by No.~1 in Table~\ref{tab:Z9_2})
over $\Z_9$ of length $16$ with Hamming weight $6$ has the following generator
matrix $G_4$:

\[  G_4 = \left [
\begin{array}{cccccccccccccccccccc}
 1\ 0\ 0\ 0\ 0\ 0\ 0\ 0\ 4\ 4\ 1\ 1\ 1\ 0\ 0\ 0\\
 0\ 1\ 0\ 0\ 0\ 0\ 0\ 0\ 7\ 2\ 7\ 2\ 0\ 1\ 0\ 0\\
 8\ 6\ 1\ 4\ 1\ 0\ 0\ 0\ 0\ 0\ 4\ 5\ 1\ 1\ 1\ 0\\
 3\ 2\ 1\ 2\ 0\ 1\ 0\ 0\ 0\ 0\ 2\ 2\ 2\ 7\ 0\ 1\\
 4\ 8\ 6\ 1\ 0\ 4\ 8\ 1\ 1\ 0\ 0\ 0\ 1\ 3\ 5\ 0\\
 2\ 8\ 2\ 6\ 7\ 6\ 8\ 2\ 0\ 1\ 0\ 0\ 3\ 8\ 0\ 4\\
 5\ 2\ 5\ 6\ 2\ 3\ 1\ 7\ 7\ 7\ 1\ 0\ 1\ 0\ 2\ 2\\
 5\ 0\ 1\ 1\ 6\ 0\ 3\ 3\ 5\ 0\ 1\ 1\ 0\ 1\ 2\ 7\\
\end{array}
\right].
\]

Using Proposition~\ref{prop:BU_2} with $G_4$, we obtain many
inequivalent self-dual codes of length $20$ with Hamming weight $6$
and distinct values of $A_6$. In Table~\ref{tab:Z9_3} we display ten
such codes, where $\tau$ denotes the kissing number of the
corresponding lattices $\Lambda(C)$. From the three distinct kissing
numbers, we know that we have constructed at least three of the $12$
inequivalent optimal Type I lattices of dimension $20$
(see~\cite[Ch.~16]{ConSlo99} or~\cite{Gab04}). It is interesting to
note that in Table~\ref{tab:Z9_3} the lattice $\Lambda(C)$ from the
$10$th code with $\tau=120$ has
$|{\mbox{Aut}}\Lambda(C)|=31310311587840$ while the others with
$\tau=120$ have $|{\mbox{Aut}}\Lambda(C)|=4299816960000$. Hence we
have constructed at least four of the $12$ inequivalent optimal Type
I lattices of dimension $20$. The first code in Table~\ref{tab:Z9_3}
has the generator matrix $G_5$ as follows:

{\small
\[  G_5 = \left [
\begin{array}{cccccccccccccccccccc}
1\ 0\ 0\ 0\ 0\ 0\ 0\ 0\ 0\ 0\ 4\ 4\ 4\ 1\ 1\ 1\ 1\ 1\ 0\ 0\\
0\ 1\ 0\ 0\ 0\ 0\ 0\ 0\ 0\ 0\ 6\ 6\ 2\ 3\ 1\ 1\ 1\ 1\ 0\ 0\\
4\ 4\ 7\ 0\ 1\ 0\ 0\ 0\ 0\ 0\ 0\ 0\ 4\ 4\ 1\ 1\ 1\ 0\ 0\ 0\\
5\ 6\ 4\ 7\ 0\ 1\ 0\ 0\ 0\ 0\ 0\ 0\ 7\ 2\ 7\ 2\ 0\ 1\ 0\ 0\\
7\ 7\ 1\ 0\ 8\ 6\ 1\ 4\ 1\ 0\ 0\ 0\ 0\ 0\ 4\ 5\ 1\ 1\ 1\ 0\\
5\ 5\ 2\ 0\ 3\ 2\ 1\ 2\ 0\ 1\ 0\ 0\ 0\ 0\ 2\ 2\ 2\ 7\ 0\ 1\\
1\ 3\ 8\ 5\ 4\ 8\ 6\ 1\ 0\ 4\ 8\ 1\ 1\ 0\ 0\ 0\ 1\ 3\ 5\ 0\\
2\ 7\ 0\ 8\ 2\ 8\ 2\ 6\ 7\ 6\ 8\ 2\ 0\ 1\ 0\ 0\ 3\ 8\ 0\ 4\\
3\ 5\ 7\ 5\ 5\ 2\ 5\ 6\ 2\ 3\ 1\ 7\ 7\ 7\ 1\ 0\ 1\ 0\ 2\ 2\\
7\ 5\ 6\ 4\ 5\ 0\ 1\ 1\ 6\ 0\ 3\ 3\ 5\ 0\ 1\ 1\ 0\ 1\ 2\ 7\\
\end{array}
\right].
\]}

Applying Proposition~\ref{prop:BU_2} to $G_5$ produces several inequivalent self-dual
codes of length $24$ with Hamming weight $6$. Their corresponding lattices $\Lambda(C)$
have minimum norm $3$, and thus each must be the {\it odd Leech lattice}.
We list three codes in Table~\ref{tab:Z9_4} where the twelve entries of the right
side of ${\bf{x}_1}$ and ${\bf{x}_2}$ respectively are written in
the second and the third column.

\begin{table}
\caption{Self-dual codes of length $12$ over GR$(3^2,1)= \Z_9$ from
$G_2$} \label{tab:Z9_1} \centering \vspace{10pt}
\begin{tabular}{|c|c|c|c|c|}
\hline Code No. & ${\bf{x}_1}=(0 0 x_3 \dots x_{8})$ &
${\bf{x}_2}=(0 0 x_3 \dots x_{8})$ & $A_6$ & $\mu(\Lambda(C))$ \\
\hline
1 & 4 5 1 1 1 0 & 2 2 2 7 0 1 & 516 & 2\\
2 & 4 5 1 1 1 0 & 8 6 5 4 1 1 & 552 & 2\\
3 & 4 5 1 1 1 0 & 5 3 2 7 1 1 & 444 & 2\\
4 & 4 5 1 1 1 0 & 8 3 8 7 1 1 & 480 & 2 \\
5 & 4 5 1 1 1 0 & 2 5 5 4 3 1 & 588 & 2 \\
6 & 4 5 1 1 1 0 & 2 2 8 6 4 1 & 408 & 2  \\
7 & 4 5 1 1 1 0 & 3 5 5 5 7 1 & 624 & 2  \\
8 & 5 5 1 1 1 0 & 0 8 7 2 5 8 & 660 & 2\\
 \hline
\end{tabular}
\end{table}

\begin{table}
\caption{Self-dual codes of length $16$ over GR$(3^2,1)= \Z_9$ from
$G_3$} \label{tab:Z9_2} \centering \vspace{10pt}
\begin{tabular}{|c|c|c|c|c|}
\hline
Code No. & ${\bf{x}_1}=(00 0 0 x_{5} \dots x_{12})$ &
${\bf{x}_2}=(0000 x_{5} \dots x_{12})$ & $A_6$ & $\mu(\Lambda(C))$ \\
\hline
1 & 4 4 1 1 1 0 0 0 & 7 2 7 2 0 1 0 0 & 266 & 2 \\
2 & 4 4 1 1 1 0 0 0 & 7 4 8 2 0 1 0 0 & 278 & 2\\
3 & 4 4 1 1 1 0 0 0 & 1 8 5 4 0 1 0 0 & 248 & 2\\
4 & 4 4 1 1 1 0 0 0 & 4 8 1 5 0 1 0 0 & 254 & 2\\
5 & 4 4 1 1 1 0 0 0 & 1 8 4 5 0 1 0 0 & 260 & 2\\
6 & 4 4 1 1 1 0 0 0 & 1 1 5 5 0 1 0 0 & 284 & 2\\
7 & 4 4 1 1 1 0 0 0 & 7 4 2 8 0 1 0 0 & 296 & 2\\
8 & 4 4 1 1 1 0 0 0 & 1 5 4 8 0 1 0 0 & 338 & 2\\
9 & 4 4 1 1 1 0 0 0 & 8 1 2 6 1 1 0 0 & 272 & 2\\
10& 4 4 1 1 1 0 0 0 & 7 5 3 1 2 1 0 0 & 242 & 2\\
11& 4 4 1 1 1 0 0 0 & 8 2 1 1 3 1 0 0 & 302 & 2\\
12& 4 4 1 1 1 0 0 0 & 2 8 1 1 3 1 0 0 & 290 & 2\\
13& 4 4 1 1 1 0 0 0 & 2 2 4 1 6 1 0 0 & 326 & 2\\
14& 4 4 1 1 1 0 0 0 & 1 7 2 5 6 1 0 0 & 230 & 2\\
15& 4 4 1 1 1 0 0 0 & 5 4 7 2 0 2 0 0 & 308 & 2\\
16& 4 4 1 1 1 0 0 0 & 8 7 8 4 0 2 0 0 & 314 & 2\\
17& 4 4 1 1 1 0 0 0 & 1 8 4 3 2 2 0 0 & 320 & 2\\
18& 4 4 1 1 1 0 0 0 & 7 2 4 0 5 2 0 0 & 332 & 2\\
19& 4 4 1 1 1 0 0 0 & 4 1 2 8 6 2 0 0 & 344 & 2\\
20& 4 4 1 1 1 0 0 0 & 7 2 3 7 8 7 0 0 & 236 & 2\\
\hline
\end{tabular}
\end{table}

\begin{table}
\caption{Self-dual codes of length $20$ over GR$(3^2,1)= \Z_9$ from
$G_4$} \label{tab:Z9_3} \centering \vspace{10pt}
\begin{tabular}{|c|c|c|c|c|c|}
                              \hline
$\#$ & ${\bf{x}_1}=(0 \cdots 0 x_{7} \dots x_{16})$ &
${\bf{x}_2}=(0 \dots 0 x_{7} \dots x_{16})$ & $A_6$, $A_7$ & $\mu(\Lambda(C))$ & $\tau$\\
\hline
1& 4 4 4 1 1 1 1 1 0 0 & 6 6 2 3 1 1 1 1 0 0 & 138, 138 &2 & 152\\
2& 4 4 4 1 1 1 1 1 0 0 & 4 4 4 1 1 1 1 1 0 0 & 138, 60& 2 & 152\\
3& 4 4 4 1 1 1 1 1 0 0 &2 5 2 5 1 1 1 1 0 0  & 138,132 & 2 & 152\\
4& 4 4 4 1 1 1 1 1 0 0 &8 5 5 5 1 1 1 1 0 0  &138, 36  & 2& 120\\
5& 4 4 4 1 1 1 1 1 0 0 &5 8 5 5 1 1 1 1 0 0  &138, 90  &2& 120\\
6&4 4 4 1 1 1 1 1 0 0  &5 5 8 5 1 1 1 1 0 0  &132, 48 &2&120\\
7&4 4 4 1 1 1 1 1 0 0  &6 2 3 6 1 1 1 1 0 0 &144, 36 &2 &120\\
8& 4 4 4 1 1 1 1 1 0 0 &5 7 2 2 2 1 1 1 0 0 &120, 30 &  2& 152\\
9&4 4 4 1 1 1 1 1 0 0  &2 6 6 1 3 1 1 1 0 0  &126, 42  & 2 & 184\\
10& 4 4 4 1 1 1 1 1 0 0 &6 5 4 6 3 1 1 1 0 0 &126, 36  & 2 &120\\
 \hline
\end{tabular}
\end{table}

\begin{table}
\caption{Self-dual codes of length $24$ over GR$(3^2,1)= \Z_9$ from
$G_5$} \label{tab:Z9_4} \centering \vspace{10pt}
\begin{tabular}{|c|c|c|c|c|}  \hline
Code No. & ${\bf{x}_1}=(0 \cdots 0 x_{9} \dots x_{24})$ &
${\bf{x}_2}=(0 \dots 0 x_{9} \dots x_{24})$& $A_6$ & $\mu(\Lambda(C))$ \\
\hline
1&4 3 2 1 1 1 1 1 1 0 0 0 & 7 7 1 4 7 2 6 1 1 0 0 0 & 48 & 3\\
2&4 3 2 1 1 1 1 1 1 0 0 0 & 2 1 2 4 7 2 6 1 1 0 0 0 &40 &3 \\
3&4 3 2 1 1 1 1 1 1 0 0 0 & 4 7 6 2 2 1 7 1 1 0 0 0 &32 &3  \\
 \hline
\end{tabular}
\end{table}


\section{Conclusion}
\label{sec:con-Building}

We have completed the open cases of the building-up construction for
self-dual codes over $GF(q)$ with $q \equiv 3 \pmod 4$ and over $\Z_{p^m}$
and Galois rings $GR(p^m, r)$ with $p \equiv 3 \pmod 4$.
We have also generalized the building-up construction for self-dual codes
to codes over finite chain rings. As a result, the building-up construction
works over any finite fields $GF(q)$, finite rings $\Z_{p^m}$, and Galois rings $GR(p^m, r)$.

We have seen that the building-up construction is a very efficient
way of finding many self-dual codes of reasonable lengths. In
particular, we construct $945$ new extremal self-dual ternary
$[32,16,9]$ codes with trivial automorphism groups, and we obtain
new optimal self-dual $[16,8,7]$ codes over $GF(7)$ and new
self-dual codes over $GF(7)$ with the best known parameters
$[24,12,9]$.
We also construct many new self-dual codes over $\mathbb Z_9$ of lengths $12, 16, 20$
all with minimum Hamming weight $6$, which is the best possible minimum Hamming weight
that free self-dual codes over $\Z_9$ of these lengths can attain. Furthermore, from the constructed
codes over $\Z_9$, we are able to reconstruct optimal Type I lattices of dimensions $12, 16, 20,$ and $24$
using Construction $A$. We conclude that our building-up construction can provide
a nice way of constructing optimal Type I lattices as well as self-dual codes.

\section*{Acknowledgment}

The authors would like to thank Dr. Thomas Feulner for his comments. J.-L. Kim was partially supported by the Project Completion Grant (year 2011-2012) at the University of Louisville.



\begin{thebibliography}{99}

\bibitem{AG} C. Aguilar Melchor and P. Gaborit,
``On the classification of extremal $[36, 18, 8]$ binary self-dual codes,'' {\em IEEE Transactions on Information Theory}, vol. 54, no 10, pp. 4743--4750, 2008.

\bibitem{AguGabKimSokSol} C. Aguilar-Melchor, P. Gaborit, J.-L. Kim, L. Sok, and P. Sol\'{e},
 Classification of extremal and $s$-extremal binary self-dual codes of length $38$, to appear in IEEE Trans. Inform. Theory.

\bibitem{BanDouHarOur99} E. Bannai, S.T. Dougherty, M. Harada, M. Oura,
{\it Type II codes, even unimodular lattices, and invariant rings},
IEEE Trans. Inform. Theory vol. {\bf 45} (1999), 1194--1205.

\bibitem{BruPle} R.A. Brualdi, V. Pless,
{\it Weight enumerators of self-dual codes}, IEEE Trans. Inform. Theory
vol. {\bf 37} (1991), 1222--1225.

\bibitem{CalSlo96} A.R. Calderbank, N.J.A. Sloane,
{\it The ternary Golay code, the integers mod $9$, and the Coxeter-Todd lattice},
IEEE Inform. Theory vol. {\bf 42} no.2 (1996), 636.

\bibitem{Mag} J. Cannon, C. Playoust, ``An Introduction to Magma,''
University of Sydney, Sydney, Australia, 1994.

\bibitem{ConSlo99} J.H. Conway, N.J.A. Sloane,
``Sphere Packing, Lattices and Groups,'' 3rd Ed., Springer-Verlag, New York, 1999.

\bibitem{Dou95} S.T. Dougherty,
{\it Shadow codes and weight enumerators}, IEEE Trans. Inform. Theory vol. {\bf 41}
no.3 (1995), 762--768.

\bibitem{DouPar} S.T. Dougherty, Y. Park,
{\it Codes over the $p$-adic integer}, Des. Codes Cryptogr. vol. {\bf 39} no 1 (2006), 65--80.

\bibitem{DouKimLiu} S.T. Dougherty, J.-L. Kim, H. Liu,
{\it Constructions of self-dual codes over chain rings}, preprint, 2008.

\bibitem{Gab04} P. Gaborit, {\it Construction of new extremal unimodular lattices},
European J. Combin. vol. {\bf 25} (2004), 549--564.

\bibitem{GulHar99} T.A. Gulliver, M. Harada,
{\it New optimal self-dual codes over $GF(7)$}, Graphs Combin. vol. {\bf 15} (1999), 175--186.

\bibitem{GulHarMiy} T.A. Gulliver, M. Harada, H. Miyabayashi,
{\it Double circulant and quasi-twisted self-dual codes over $\mathbb
F_5$ and $\mathbb F_7$}, Adv. Math. Commun. {\bf 1} No. 2 (2007), 223--238.

\bibitem{GulKimLee} T.A. Gulliver, J.-L. Kim, Y. Lee,
{\it New MDS or near-MDS self-dual codes}, IEEE Trans. Inform. Theory, {\bf 54} no.9 (2008), 4354--4360.

\bibitem{Hensel1}  F.Q. Gouv\^ea,
``$P$-adic numbers: an introduction,'' Springer-Verlag, New York-Berlin, 1993.

\bibitem{GreVit99} M. Greferath, E. Viterbo,
{\it On $\mathbb Z_4$- and $\mathbb Z_9$-linear lifts of the Golay codes},
IEEE Trans. Inform. Theory, {\bf 45} no.7 (1999), 2524--2527.

\bibitem{Har_09} M. Harada, personal communication on April 25, 2009.

\bibitem{Har97} M. Harada, {\it The existence of a self-dual $[70, 35, 12]$ code
and formally self-dual codes}, Finite Fields Appl. {\bf  3} (1997), 131--139.

\bibitem{Har_98} M. Harada, {\it New extremal ternary self-dual codes},
Austral. J. Combin. {\bf 17} (1998), 133--145.

\bibitem{Har_01} M. Harada, {\it An extremal ternary self-dual
$[28, 14, 9]$ code with a trivial automorphism group}, Discrete Math. {\bf 239} (2001), 121--125.

\bibitem{HarHozKhaKho} M. Harada, W. Holzmann, H. Kharaghani,
M. Khorvash, {\it Extremal ternary self-dual codes constructed from negacirculant matrices}, Graphs Combin. {\bf 23} (2007), 401--417.

\bibitem{HarKha} M. Harada, H. Kharaghani,
{\it Orthogonal designs, self-dual codes, and Leech lattice},
J. Combin. Designs, {\bf 13} (2005), 184--194.

\bibitem{HarMunVen} M. Harada, A. Munemasa, B. Venkov,
{\it Classification of ternary extremal self-dual codes of length 28},
to apprear in Math. Comp.

\bibitem{HarOst} M. Harada, P.R.J. \"Osterg{\aa}rd,
{\it Self-dual and maximal self-orthogonal codes over $\mathbb F_7$},
Discrete Math. {\bf 256} (2002), 471--477.

\bibitem{Huf_92} W.C. Huffman, {\it On extremal self-dual ternary codes
 of lengths 28 to 40}, IEEE Trans. Inform. Theory, {\bf 38} (1992), 139--400.

\bibitem{Huf_05} W.C. Huffman, {\it On the classification and enumeration
of self-dual codes}, Finite Fields Appl. {\bf 11} (2005), 451--490.

\bibitem{HufPle} W.C. Huffman, V. S. Pless,
``Fundamentals of Error-correcting Codes,'' Cambridge: Cambridge
University Press, 2003.

\bibitem{IreRos} K.F. Ireland, M. Rosen,
``A classical introduction to modern number theory,''
Springer-Verlag, New York-Berlin, 1982.

\bibitem{Kim} J.-L. Kim,
{\it New extremal self-dual codes of lengths $36, 38$ and $58$}, IEEE Trans.
Inform. Theory, {\bf 47} (2001), 386--393.

\bibitem{KimLee} J.-L. Kim, Y. Lee,
{\it Euclidean and Hermitian self-dual MDS codes over large finite
fields}, J. Combin. Theory Ser. A, {\bf 105} (2004), 79--95.

\bibitem{KimLee2} J.-L. Kim, Y. Lee,
{\it Construction of MDS Self-dual codes over Galois rings}, Des. Codes Cryptogr.,
{\bf 45} no. 2 (2007), 247--258.

\bibitem{Koc89} H. Koch,
{\it On self-dual doubly-even codes of length 32}, J. Combin. Theory Ser. A, {\bf 51} (1989), 63--76.

\bibitem{LeeLee} H. Lee, Y. Lee,
{\it Construction of self-dual codes over finite rings ${\mathbb
Z}_{p^m}$}, J. Combin. Theory Ser. A, {\bf 115} (2008), 407--422.

\bibitem{LeoPleSlo} J. S. Leon, V. Pless, N. J. A. Sloane,
{\it On ternary self-dual codes of length 24}, IEEE Trans. Inform. Theory, {\bf 27} no. 2 (1981), 176--180.

\bibitem{Ana} G.~H. Norton, A. S\u al\u agean,
{\it On the Hamming  distance of linear codes over a finite chain ring},
IEEE Trans. Inform. Theory, {\bf 46}, no. 3 (2000), 1060--1067.

\bibitem{Ple75} V. Pless, {\it On the classification and enumeration of
self-dual codes}, J. Combin. Theory Ser. A, {\bf 18} no. 3 (1975),
313--335.

\bibitem{Ple72} V. Pless, {\it Symmetry codes over GF($3$) and new five-designs},
J. Combin. Theory Ser. A, {\bf 12} no. 1 (1972), 119--142.

\bibitem{PleTon} V. Pless, V. Tonchev,
{\it Self-dual codes over $GF(7)$}, IEEE Trans. Inform. Theory {\bf 33}
no. 5 (1987), 723--727.

\bibitem{RaiSlo} E. Rains, N.J.A. Sloane,
{\it Self-dual codes}, in: V.S. Pless, W.C. Huffman (Eds.), ``Handbook of
Coding Theory,'' Elsevier, Amsterdam. The Netherlands, 1998.

\end{thebibliography}
\end{document}